# It's only fair when I think it's fair: How Gender Bias Alignment Undermines Distributive Fairness in Human-AI Collaboration

Domenique Zipperling*
University of Bayreuth & Fraunhofer FIT
Bayreuth, Germany
domenique.zipperling@uni-bayreuth.de

Luca Deck*
University of Bayreuth & FIM Research Center
Bayreuth, Germany
luca.deck@fim-rc.de

Julia Lanzl[†]
University of Hohenheim
Hohenheim, Germany
Fraunhofer FIT
Augsburg, Germany
julia.lanzl@uni-hohenheim.de

Niklas Kühl[†]
University of Bayreuth & Fraunhofer FIT
Bayreuth, Germany
niklas.kuehl@fit.fraunhofer.de

## Abstract

Human-AI collaboration is increasingly relevant in consequential areas where AI recommendations support human discretion. However, human-AI teams' effectiveness, capability, and fairness highly depend on human perceptions of AI. Positive fairness perceptions have been shown to foster trust and acceptance of AI recommendations. Yet, work on confirmation bias highlights that humans selectively adhere to AI recommendations that align with their expectations and beliefs—despite not being necessarily correct or fair. This raises the question whether confirmation bias also transfers to the alignment of gender bias between human and AI decisions. In our study, we examine how gender bias alignment influences fairness perceptions and reliance. The results of a 2x2 between-subject study highlight the connection between gender bias alignment, fairness perceptions, and reliance, demonstrating that merely constructing a "formally fair" AI system is insufficient for optimal human-AI collaboration; ultimately, AI recommendations will likely be overridden if biases do not align.

---

*Authors contributed equally to this research.
[†]Authors supervised this research equally.

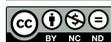

## 1 Introduction

Artificial Intelligence (AI) is increasingly used to support human decision-making by providing recommendations in consequential areas, such as credit scoring [2, 53] or hiring [62]. To mitigate potentially harmful disparities for protected groups, the discipline of fair machine learning (ML) [57, 90] has produced a broad set of fairness-enhancing techniques to enforce fairness metrics in system outputs. However, where AI does not operate autonomously, it is vital not to consider system outputs isolated from human discretion. Therefore, recent studies emphasize the need to promote not only fair predictions but, ultimately, fair decisions [66, 84].

Against this backdrop, behavioral studies suggest that reliance behavior is strongly linked to human perceptions such as fairness and trustworthiness [50, 73, 75]. Fairness perceptions are, in turn, influenced by personal traits such as age, gender, or political views [35, 39, 63]. Despite a better understanding of fairness perceptions and reliance behavior, achieving fair and effective human-AI collaboration remains challenging, as AI is susceptible to historical biases, and humans are prone to a variety of cognitive biases [44, 57]. To shed light on the intricate interconnection of algorithmic and human biases occurring in human-AI collaboration, prior work has studied the role of explanations [70] and confirmation bias [7, 76]. Confirmation bias is a particularly relevant concept as it describes the tendency of humans to favor information that aligns with their pre-existing beliefs [91] and selectively adhere to AI recommendations that align with their expectations [3]. For example, Wang et al. [89] find that students prefer a biased career recommendation system over a debiased one, largely due to underlying gender biases. Similarly, Baumler and Daumé [8] show that language model suggestions aligned with common social stereotypes are more likely to be accepted by human authors, while anti-stereotypical suggestions have only a limited impact on reducing bias in generated stories.

*Our work.* In this work, we study the alignment between AI recommendations and human expectations concerning fairness. Specifically, we examine how an alignment influences *(i)* fairness perceptions of AI, *(ii)* the reliance of human decision-makers interacting with AI recommendations ("human-AI teams"), and *(iii)* the distributive fairness of the final decisions. As a specific example of fairness, we focus on *gender bias*, defined as disparities in acceptance rates between genders—while acknowledging the existence of various fairness metrics and sensitive attributes [57]. Based on this definition, *gender bias alignment* quantifies the similarity of gender bias between human decisions and AI recommendations. Therefore, we raise the following research questions:

**RQ 1.** *How does the alignment of gender bias between humans and AI influence human fairness perception of AI?*

**RQ 2.** *How does the alignment of gender bias between humans and AI influence reliance behavior in human-AI teams?*

**RQ 3.** *How does the alignment of gender bias between humans and AI influence the outcome of human-AI teams w.r.t to distributive fairness?*

To answer these questions, we conduct two 2x2 between-subject experiments in which humans are teamed up with AI to complete classification tasks in high-stakes contexts. As treatments, we introduce two degrees of gender bias in the AI recommendations and augment the recommendations with explanations in one group. In experiment I, participants iteratively perform credit scoring for ten applicant profiles with the support of AI—either with or without gender bias. For each applicant, participants initially decide on the eligibility for credit before receiving an AI recommendation—some participants are additionally supported with explanations regarding the recommendation—and then make a final decision. At the end of the study, participants report their perceived fairness. Experiment II follows the same procedure in a different context. Here, participants are asked to judge a criminal's probability of reoffending within the next two years.

*Findings and implications. First*, we observe a significant influence of gender bias alignment on participants' fairness perceptions. The more gender bias aligns, the fairer the AI is perceived. *Second*, we confirm prior findings (e.g., [70]) showing that fairness perceptions can influence reliance behavior in human-AI teams. Humans tend to override the AI's recommendation less if the AI is perceived as fairer. *Third*, gender bias alignment influences reliance behavior as a higher alignment reduces the amount of overridden AI recommendations. *Fourth*, the distributive fairness of human-AI team decisions strongly depends on gender bias alignment. If gender bias does not align, the AI's influence on distributive fairness diminishes. As humans tend to override recommendations more often, formally fair recommendations can be turned into formally unfair decisions. *Fifth*, we do not observe a significant influence of explanations on participants' fairness perceptions, thereby adding to the inconclusive findings of prior research [10, 24, 74, 76].

Overall, our findings have major implications for the effectiveness of fairness-enhancing techniques as well as the design of human-AI teams. Our study reveals that not only fairness perceptions but also reliance behavior are significantly tied to the alignment of human and AI biases (such as gender bias). Alignment of model properties and human values requires heightened attention [60]. If this alignment is disregarded, human decision-makers are likely to override potentially fair recommendations, rendering fairness-enhancing techniques ineffective. Our study, therefore, reinforces recent calls to shift the focus from (formally) fair AI recommendations to ultimately *fair decisions of human-AI teams*.

## 2 Background

We start by summarizing the predictors of reliance behavior and fairness perceptions studied in prior works. Studies on AI reliance and fairness have thus far primarily focused on human and algorithmic predictors independently [81]. This leaves a research gap concerning the *alignment* of characteristics present in humans and AI, such as gender bias. We introduce the two overarching themes of AI reliance and fairness and relate them to two central concepts: explainable AI and confirmation bias.

### 2.1 Human-AI Collaboration and Reliance

In recent years, significant progress has been made in AI technology, and human-AI teams have found their way into many high-stakes domains, such as medicine and recruiting [52, 62, 82]. Collaboration between human decision-makers and AI can ideally lead to increased performance, leveraging the complementary strengths of both sides [5, 67]. Humans tend to excel at tasks involving abstract thinking, like generalization and intuition [41], whereas AI excels in analytic, data-driven tasks such as recognizing complex patterns and efficient probabilistic computations [21, 45]. Despite these strengths, human decision-making is susceptible to cognitive biases, which can distort judgment and affect outcomes [86], translating to human-AI teams [44]. In the same way, AI systems can be influenced by human biases, as they can inherit them from human-labeled training data or human influence in system development [37, 87].

In human-AI team settings, novel challenges arise. The effectiveness of collaboration and task completion between human decision-makers and AI can be broadly divided into three patterns. When confronted with an AI recommendation, human decision-makers can *over-rely*, often stemming from a phenomenon labeled "automation bias" [3, 18, 55, 69], *under-rely*, often stemming from a phenomenon labeled "algorithm aversion" [22], or *appropriately rely* [50, 68]. Appropriate reliance describes the amount of human reliance on AI recommendations that lead to an overall better performance [68]—assuming each party can contribute something to the overall performance [40]. Research on human-AI collaboration and appropriate reliance has long only considered traditional performance metrics such as accuracy. However, recent publications have proposed to take fairness considerations into account [70, 71].

### 2.2 AI Fairness

AI fairness becomes an important consideration, extending traditional performance metrics in sensitive and high-stakes domains involving human fates such as credit scoring [2, 53] or hiring [62]. Although there is no universal definition of what constitutes a fair decision, AI fairness can be broadly defined as the absence of discriminatory or unjust consequences of algorithmic decisions [92].

Distributive fairness in AI is often captured as a formal model property in the form of demographic parity (DP), but also as a psychological construct in the form of fairness perceptions. Formally, fairness can be evaluated in the form of metrics capturing disparities between socially relevant demographic categories, such as gender or race. For example, DP is a popular fairness metric that measures the acceptance rate among groups, where equal acceptance rates are considered fair [14]. On the other hand, fairness perceptions are subjective human assessments of aspects within a decision context where distributive fairness perceptions are primarily concerned with outcomes [17, 38, 42]. Starke et al. [81] review 58 empirical studies and summarize a range of significant predictors influencing fairness perceptions: Algorithmic predictors include characteristics of the model design, such as formal fairness properties or supplementary explanations, whereas human predictors capture individual human factors such as AI literacy or political views. Their findings suggest that both the design of the AI system, e.g., with regard to distributive fairness, as well as human preferences for distributive fairness, are crucial predictors. Despite that, research on the explicit role of alignment between distributive fairness and fairness perceptions is scarce.

Recent research indicates a connection between perceived fairness and reliance behavior. Schoeffer et al. [70] demonstrate a negative correlation between perceived fairness and the likelihood of overriding an AI recommendation. They find that the fairer an AI is perceived, the less likely humans are to override the AI recommendation, emphasizing the need to understand factors that influence fairness perception. The following sections explore the role of explainable AI and confirmation bias concerning reliance behavior and fairness perceptions.

### 2.3 Explainable AI

The field of explainable AI (XAI) aims to mitigate challenges arising from black-box behavior and to increase the transparency of AI-informed decision processes. Broadly speaking, there are two common approaches to make AI recommendations more comprehensible to humans: *Local* interpretability provides explanations for individual model outputs, while *global* interpretability attempts to explain the general behavior of an AI system [1, 79]. A different approach to classifying XAI methods is to divide them into model-specific or intrinsic and model-agnostic methods, where models are either built around explainability or, providing more flexibility, explanation methods are applied post hoc to predictions [1, 79].

*Effects on fairness perceptions.* The influence of explanations on fairness perceptions is well reported. Still, it remains inconclusive, as some studies demonstrated that presenting explanations can enhance fairness perceptions [74, 76, 77], while other studies stress that the effect is highly context-dependent and explanations can even decrease perceived fairness [10, 51]. For example, the style (e.g., case-based vs. demographic-based) and level (local vs. global) are important algorithmic predictors for perceived fairness [10]. Dodge et al. [24] note that global explanations can enhance fairness perceptions, whereas local explanations rather help to identify fairness discrepancies between instances. Recent studies have also explored how local explanations impact fairness perceptions and collaboration between humans and AI. Highlighting protected attributes such as gender in explanations was found to impact fairness perceptions and human-AI collaboration [70]. Crucially, explanations can also be misused to manipulate certain stakeholders by building unwarranted trust [26, 46] or fooling affected parties into believing an AI system is fair [19, 49].

Apart from the style of explanation, it is crucial to account for the model properties the explanations are based on, i.e., whether they reveal varying degrees of fairness. To the best of our knowledge, only two studies introduce different levels of formal fairness [4, 51]. Both find that distributive fairness significantly moderates the way explanations alter fairness perceptions.

*Effects on reliance.* Explanations are expected to increase transparency, acceptance, and appropriate reliance [78, 85]. However, the relationship between XAI, human perceptions, and reliance behavior is not straightforward and prone to a range of undesirable side effects, such as unwarranted trust [46] or reinforcement of stereotypes [70]. While explanations are often hoped to promote reliance on AI systems [64], empirical evidence on their usefulness remains inconclusive. Recent studies suggest that explanations can sometimes harm reliance. For instance, explanations can increase the likelihood of users accepting AI recommendations [6] but can also lead to issues of over-reliance [12]. Focusing on explanations highlighting gender- or task-specific sections of applicant profiles, Schoeffer et al. [70] find that feature-based explanations do not affect the likelihood of overriding correct or incorrect AI recommendations. However, when only task-relevant words are highlighted, reliance behavior reinforces stereotypes, potentially reducing the distributive fairness of AI-informed decisions.

In our study, we vary the AI systems' formal fairness and provide global explanations to the human decision-makers to understand how additional information about the AI system beyond mere recommendations shapes fairness perceptions and reliance behavior. Moreover, we do not only account for the AI but also for human characteristics in the form of confirmation bias.

### 2.4 Confirmation Bias

Confirmation bias, central to human-AI team research, describes how humans' trust in AI recommendations aligns with their beliefs [91]. Similar dynamics appear in psychology, shaping theories like *similarity-attraction theory* [13], which posits that shared traits, such as demographics, personality, and values, foster attraction [65]. *Homophily theory*, particularly *value homophily*, extends this, explaining how shared values and beliefs drive social connections [56]. *Cognitive dissonance theory* [30] further suggests that humans seek psychological consistency, with belief-disconfirmation leading to misperception, misinterpretation, or rejection of conflicting information [36].

*Effects on reliance and fairness perceptions.* Confirmation bias can strongly impact the effectiveness of human-AI teams. The term "selective adherence" [3] connects AI reliance with confirmation bias. Bashkirova and Krpan [7] investigate the impact of confirmation bias in a clinical context. The results indicate that alignment of AI recommendations and previous beliefs leads to higher perceived trustworthiness and acceptance. They find a correlation between professional knowledge and automation bias when AI decisions match experts' presumptions. Similar effects are reported in the interaction of street-level bureaucrats with AI recommendations [72]. Additionally, in the context of organ donations, people are found to rely more strongly on AI recommendations when they perceive the AI system's values to be similar to their own [60]. Applying these findings on confirmation bias to fairness perceptions gives rise to a range of intriguing hypotheses, which we derive in the subsequent section.

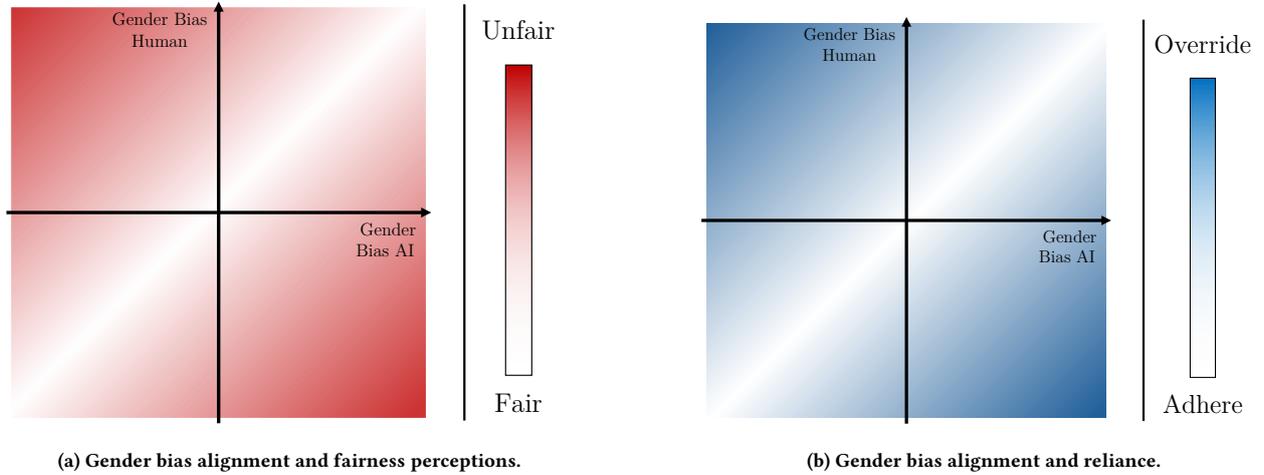

(a) Gender bias alignment and fairness perceptions.

(b) Gender bias alignment and reliance.

Figure 1: Visual illustrations for the central hypotheses on gender bias alignment effects on fairness perceptions and reliance. The axes represent the ratio of acceptance rates, indicating differences between men and women for AI recommendations (x-axis) vs. human decisions (y-axis). Smaller values signify demographic parity, i.e., equal acceptance rates for men and women. In areas of high gender bias alignment (white), fairness perceptions are favorable (a) and overrides are minimal (b). Greater misalignment (shaded areas) corresponds to worse fairness perceptions (a) and more frequent overrides (b). Notably, positive fairness perceptions require *alignment* of gender biases between AI and humans, not necessarily small AI bias.

## 3 Theoretical Development

In this section, we combine the theories and findings above into hypotheses that guide our research design. We illustrate our underlying reasoning abstractly and visually (see Figure 1), conceptualize our research model (Figure 2), state our research hypotheses, and describe our measurement concepts.

### 3.1 Development of Hypotheses

Our study takes a novel perspective on fairness perceptions and reliance behavior by linking the gender bias of humans and AI (RQ 1). We investigate the relationship between fairness perceptions and reliance (RQ 2) and the effect on distributive fairness (RQ 3). Figure 1 illustrates our hypotheses: human and AI gender bias, quantified by differences in acceptance rates (e.g., in loan applications), influence perceptions and reliance. Larger absolute differences indicate higher gender bias. We propose that fairness perceptions (Figure 1a) and reliance (Figure 1b) depend on gender bias alignment. With high alignment (white area), fairness perceptions are more favorable, and overrides are minimal. Low alignment (shaded area) worsens fairness perceptions and increases overrides. Crucially, positive fairness perceptions require *bias alignment* between AI and humans, not necessarily low AI bias.

*Gender bias alignment and fairness perceptions.* A range of factors is known to impact fairness perceptions of AI systems [81]. Considering the effects of confirmation bias [7, 60], its role in shaping perceptions, and key psychological theories [13, 65, 91], we argue that the alignment in gender bias presents another crucial influence on fairness perceptions. Additionally, recent research indicates a link between explanations and fairness perceptions [10, 51, 74, 76].

While local feature-based explanations have been shown to influence fairness perceptions in human-AI teams [70], the evidence on the influence of global explanations is still scarce. Based on this, we propose the following hypotheses:

**H 1.** *The alignment in gender bias between humans and AI influences human fairness perceptions of AI.*

**H 2.** *Providing global feature importance influences human fairness perceptions of AI.*

*Gender bias alignment and reliance.* Confirmation bias and explanations influence not only perceptions but also reliance [6, 7, 70, 78]. Moreover, fairness perceptions can directly affect reliance [70]. Therefore, we formulate the following hypotheses:

**H 3.** *The alignment in gender bias between humans and AI influences reliance on AI recommendations.*

**H 4.** *Providing global feature importance influences reliance on AI recommendations.*

**H 5.** *The fairer humans perceive AI recommendations, the fewer decisions they tend to override.*

*Gender bias alignment and distributive fairness.* An effect on reliance behavior does not necessarily impact human-AI team accuracy, as errors may shift [70], with improvements of humans in some areas (e.g., more true positives) offset by declines in others (e.g., fewer true negatives). Notably, accuracy is not the only objective in human-AI teams if we adopt a more holistic perspective on AI-informed decision-making [71]. In this work, we focus on fairness-centered reliance behavior via acceptance rates, excluding accuracy considerations. Drawing on research on confirmation bias

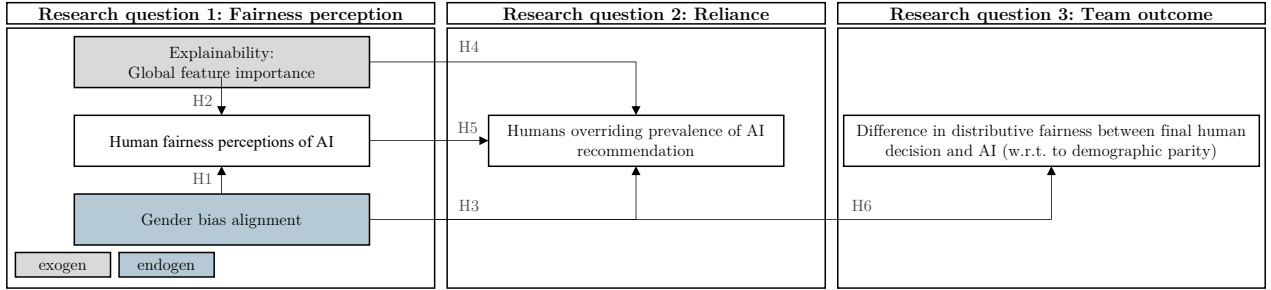

**Figure 2: Research model considering the three research questions: How does the alignment of gender bias alignment between humans and AI influence fairness perception of AI (RQ 1), reliance behavior (RQ 2), and distributive fairness (RQ 3)?**

and reliance, we hypothesize that humans are more likely to override AI recommendations when gender bias alignment is low [7, 60]. Therefore, we expect that humans will enforce their initial gender bias on the final decision, thus reducing the AI's impact on distributive fairness. As a result, the disparity in distributive fairness between the AI and the final human decision is expected to increase. To explore these dynamics, we analyze the following hypothesis:

**H 6.** *The alignment in gender bias between humans and AI influences the difference in distributive fairness of the AI and the final human decision.*

### 3.2 Measuring Concept

To address the hypotheses, we introduce the key concepts and measures for confirmation bias, reliance, and distributive fairness. We only provide the most central equation below. An account of all equations is provided in Appendix A.

*Measuring confirmation bias.* Confirmation bias in AI refers to how human expectations influence the perception and acceptance of AI systems and outputs [7]. Beyond superficial expertise alignment, we focus on value alignment in behavior, particularly gender bias—a persistent issue in historical data used for hiring and credit scoring [15, 34]. We measure gender bias alignment using acceptance rate and distributive fairness based on demographic parity. Acceptance rate (AR) is defined as the conditional probability of acceptance ($\hat{Y}$ = accepted) based on gender ($men(M), women(W)$) and decision-maker ($DM \in \{AI, human(H)\}$). AR alignment quantifies the absolute differences in AR across genders, normalized to [0, 1], with higher values indicating greater alignment (see Equation (1)). Demographic parity (DP) is the difference in acceptance probabilities for men and women by a decision-maker ($DM$), with zero indicating no bias. DP alignment is the normalized absolute difference between $DP(AI)$ and $DP(H)$ (see Equation (2)).

$$\text{Alignment}(AR) = 0.5 \cdot \big(2 - (|AR(AI, M) - AR(H, M)| \\ + |AR(AI, W) - AR(H, W)|)\big), \quad (1)$$
$$\text{Alignment}(AR) \in [0, 1]$$

$$\text{Alignment}(DP) = 0.5 \cdot (2 - |DP(AI) - DP(H)|), \quad (2)$$
$$\text{Alignment}(DP) \in [0, 1]$$

*Measuring reliance.* Reliance describes how humans respond to AI recommendations, incorporating automation bias, algorithmic aversion, and appropriate reliance [18, 22, 50, 69]. The literature on AI reliance distinguishes between one-stage and two-stage measurement approaches [25]. We adopt a two-stage design to capture participants' initial biases before AI exposure, enabling a clean analysis of human-AI bias alignment. To measure reliance, we use agreement percentage between final decisions and AI recommendations, as it provides a simple, established metric [70], ensuring comparability with prior work while maintaining a rigorous separation between initial judgment and AI-influenced outcomes.

*Measuring difference in distributive fairness.* Distributive fairness uses metrics to translate social values into computational terms [57]. While these metrics approximate fairness and involve trade-offs [31, 43], we chose DP for its simplicity and accessibility [80]. We aim to examine the relationship between gender bias alignment, the team's distributive fairness—measured by the distributive fairness of the final human decisions—and the distributive fairness of AI recommendations. To achieve this, we quantify the difference in distributive fairness as the disparity in DP between the AI recommendations and the final human decisions.

## 4 Study Design

Our study examines how gender bias alignment impacts fairness perceptions, reliance, and distributive fairness. We also assess how global explanations, such as feature importance, affect perceptions of fairness and overriding behavior. Additionally, we explore how fairness perceptions influence reliance. We conduct two 2x2 between-subject experiments, each involving a human-AI team completing a unique task: credit scoring in the first experiment and parole decisions in the second. This section details the datasets and tasks, the data processing methods used to create AI models with varying gender bias, the study design, and the data collection process. Below, we provide a general overview of both tasks, their respective data sets, and the overall experimental design. All preparatory steps—pre-processing, AI development, explanations, and experiment sample generation—are explained using the credit scoring task (Task I) as a representative example. A detailed description of the parole decision task (Task II) is included in Appendix B. Demographic information about the participants is reported for both tasks.

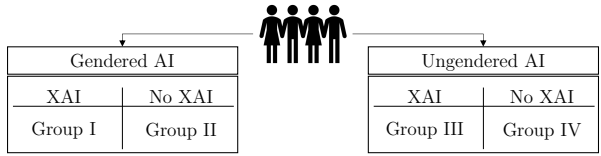

**Figure 3: Assignment of participants to treatments. Based on the 2x2 between-subjects design, participants of one task are randomly divided into four different groups. Each group has different characteristics: seeing recommendations from the gendered or ungendered AI and seeing additional explanations or not.**

## 4.1 Datasets and Tasks

We conduct two distinct experiments, each involving a different task. After introducing both tasks, we describe the preparation steps illustratively for one. In the first experiment (Task I: Credit Scoring), participants act as bank employees tasked with assigning applicants to either *eligible* or *not eligible* for credit. The task uses the US Census dataset [9], widely used in fair ML research due to its sensitive yet task-irrelevant features [29]. Since the dataset lacks explicit eligibility labels, the target variable "income" serves as a proxy, with individuals earning $50,000 or more classified as eligible. In the second experiment (Task II: Parole Decision), participants act as parole officers reviewing convict profiles to predict reoffense within two years (*will reoffend* or *will not reoffend*). The task uses the COMPAS dataset, also widely used in fair ML research [29].

## 4.2 Pre-processing, AI Development, and Explanations

To analyze gender bias alignment, we train two AI models: a gendered and an ungendered version. The ungendered model has no gender bias (measured by DP). Data pre-processing involves two steps: a shared first step and a model-specific second step. In step 1, we remove other sensitive features (ethnicity and native country). Applicants over 70 are excluded, as banks often deny credit based on age. The "capital" feature is derived from "capital gain" and "capital loss," and hard-to-interpret features are dropped. The final dataset includes Age, Capital, Current occupation, Highest education level, Hours worked per week, Marital status, Gender, and Workclass. "Hours worked per week" is grouped into three categories: below 30, 30-40, and above 40 hours. Missing values are imputed with the median, numeric features are standardized, and categorical features are one-hot encoded.

In step 2, we adjust the bias of the AI to create two datasets: a gendered and an ungendered one. For the gendered dataset, data points of a certain group are undersampled or oversampled. Specifically, applicants identifying as men who are not eligible for credit are dropped to induce a connection between eligibility and men. For the ungendered dataset, data points of applicants identifying as women who are not eligible for credit are dropped to reduce bias. After pre-processing, an 80:20 train-test split is performed, and an XGB classifier is trained on each dataset, enabling direct extraction of global explanations. The ungendered model achieves an accuracy of 85% and a distributive fairness of 0.02, indicating minimal gender bias. The gendered model achieves an accuracy of 89% and a distributive fairness of 0.7, indicating significant gender bias. A visual illustration can be found in Figure B.1.

## 4.3 Experimental Design

We conduct two 2x2 between-subject studies based on two binary classification tasks. Both tasks involve classifying people into two groups, with one group considered favorable (e.g., eligible for credit). Participants are randomly assigned to one of four groups, defined by whether they work with a gendered or ungendered AI and whether AI recommendations include explanations. The possible groups for both tasks are shown in Figure 3. Apart from these characteristics, the experiment follows the same procedure across groups, as illustrated in Figure 4. Participants are shown eleven profiles, including one attention check, resulting in ten profiles for evaluation. After reviewing each profile, participants assign a class (e.g., eligible) and report their confidence on a 5-point Likert scale. They are then shown an AI recommendation, with groups I and III additionally receiving explanations. Participants reassess the profile, assign a class, and report their confidence again. At the end, participants rate the perceived fairness of the AI system on a 5-point Likert scale and provide information about their decision-making strategies and demographic data.

## 4.4 Sample Generation

A key aspect of the study involves selecting the set of profiles that are shown to the participants. Therefore, we construct an ungendered and gendered experiment sample—comprising applicant profiles with AI recommendations—to represent either the ungendered or the gendered AI model. These samples are designed to meet the desired accuracy and distributive fairness while maintaining similar attributes to ensure comparability. First, the general characteristics of both experiment samples are defined, setting an 80% accuracy to include both error types. Gender bias, measured by DP, is established for each experiment sample independently. The ungendered experiment sample has zero parity with a 40% AR, while the gendered one has a 0.4 parity, with ARs of 60% for men and 20% for women. Both samples maintain a 50:50 men-to-women ratio. To generate experiment samples with the defined characteristics, the profiles contained in the ungendered test dataset are divided into eight groups based on gender, credit eligibility, and the AI's recommendation. Group frequencies are determined by the accuracy and parity settings. Samples are randomly drawn to ensure comparability while maintaining randomness. Further details are provided in Appendix B.2.

## 4.5 Data Collection

Our study received approval from the institutional ethics committee. Participants are recruited via Prolific, an online research platform [61]. Participants are required to be at least 18 years old and fluent in English. A total of 280 laypeople are recruited, with 140 participants assigned to each task.

For task I (Credit Scoring), 36% of participants are aged 25–34, 31% are between 35–44, 27% are between 45–54, and 6% are 55 or older. 58% of participants identify as women, 40% as men, and 2% identify as neither. Regarding AI literacy, 50% report little knowledge, 35% report moderate knowledge, 14% report some knowledge, and 1% do not report their AI knowledge. Ethnically, 64% identify as White or Caucasian, 22% as Black or African American, 10% as Asian, and 4% as other.

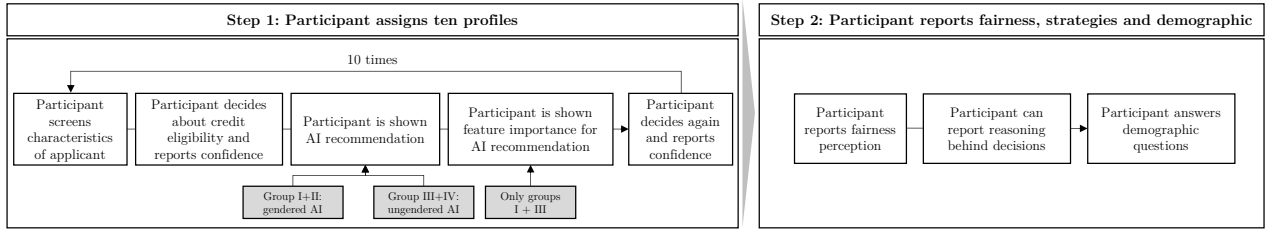

Figure 4: Two steps of the experimental setup. The experiment consists of two steps, illustrated based on task I. (1) Participants screen applicants, decide about eligibility, are supported by an AI recommendation, some are shown feature importance, and make a final decision. (2) Reporting perceived fairness and additional information.

For task II (Parole Decision), 41% of participants are aged 25–34, 31% are between 35–44, 21% are between 45–54, and 7% are 55 or older. 62% participants identify as women, 36% as men, and 2% identify as neither. Regarding AI literacy, 50% report little knowledge, 30% report moderate knowledge, 19% report some knowledge, and 1% do not report their AI knowledge. Ethnically, 68% identify as White or Caucasian, 18% as Black or African American, 9% as Asian, 4% as other, and 1% do not report their ethnicity.

## 5 Data Analysis and Results

For each research question, we analyze the data utilizing different regression methods. Before analyzing the data, pre-processing is necessary. Due to the imbalanced distribution across fairness perceptions and AI literacy, we pool groups as follows. For AI literacy, we pool groups "No Knowledge" and "Little Knowledge" as well as "Moderate Knowledge" and "A lot of Knowledge" into three levels (little/some/moderate). Similarly, we pool responses for the fairness perceptions: "Unfair" and "Slightly Unfair" as well as "Slightly Fair" and "Fair" are pooled respectively, resulting in three levels (unfair/neutral/fair). Furthermore, we remove data from participants who did not answer the necessary questions about their demographics, as we are interested in controlling for certain characteristics like gender. Pre-processing results in final sample sizes of 134 for task I and 132 for task II. After preprocessing, and given that human-AI bias alignment is central to our study, we examine the distribution of human biases to ensure broad coverage. Figure 5 shows that, for the credit scoring task, biases are approximately bell-shaped across the full range (−1 to 1), whereas for the parole decision task, biases are slightly skewed toward favoring predictions of male recidivism.

A summary of our results is provided in Table 4 at the end of the section, illustrating which factors influence fairness perception, reliance, and the team's distributive fairness for each task.

### 5.1 Gender Bias Alignment and Fairness Perceptions

As fairness perception is an ordinal variable, we utilize a proportional ordered logistic regression (POLR) to analyze the influence of the alignment and other factors on perception. POLRs are a common method to analyze ordinal-scaled variables in social science [33] and have advantages over other regression models [54]. To ensure the validity of the analysis, we utilize a Brant as well as a variance inflation factor test to assess the proportional odds assumption and multicollinearity, respectively. Based on the results and the fact that not all fairness groups are present for all options of control variables, we analyze the influence of AR alignment, alignment in distributive fairness, and the presence of XAI on the perception of

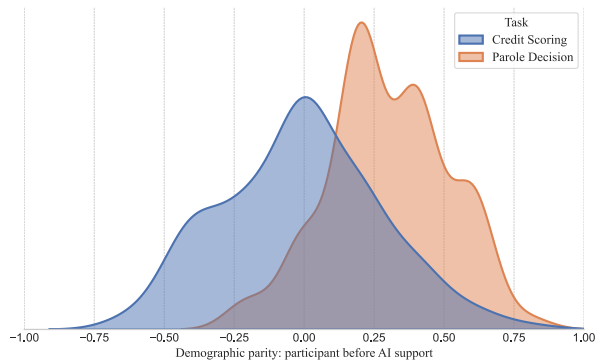

Figure 5: Distribution of the human gender bias before seeing the AI support for both tasks.

Table 1: POLR results with standard errors (SE) for fairness perception in two tasks: credit scoring vs. parole decision. This table presents the coefficients and SEs of predictors in both tasks, indicating their effect on fairness perception.

| *Dependent Variable: Fairness Perception* | | | | |
|---|---|---|---|---|
| | **Credit Scoring** | | **Parole Decision** | |
| | Coeff. | SE | Coeff. | SE |
| Alignment: Distributive Fairness | 1.290 | (1.544) | 3.444** | (1.651) |
| Alignment: Acceptance Rates | 3.206** | (1.495) | 10.380*** | (3.109) |
| Explanations Provided | 0.120 | (0.346) | -0.067 | (0.484) |
| Observations | 134 | | 132 | |

*Note:* $^{*}p<0.05$; $^{**}p<0.01$; $^{***}p<0.001$

fairness while controlling for gender and AI literacy only. We will use the same set of control variables for all subsequent analyses as well. As we want to focus on the central hypotheses of your study, the reported table will not contain the control variables.

The results for both tasks are illustrated in Table 1 and demonstrate a significant relationship with AR alignment. The higher the AR alignment, the more likely the model is to be perceived as fair. Additionally, in the second task, alignment in distributive fairness significantly influences fairness perception. Similarly, the higher DP alignment, the fairer the model is perceived. Both findings support H 1 and the influence of gender bias alignment on fairness perception. For both tasks, providing explanations in the form of global feature importance does not have a significant effect on fairness perception, thereby not supporting H 2.

### 5.2 Gender Bias Alignment and Reliance

Regarding reliance, we are interested in influences on the prevalence of overriding an AI recommendation. Therefore, we utilize a beta regression as it is well-suited for dependent variables between zero and one. Again, we use gender and AI literacy as control variables. The central results are demonstrated in Table 2 and show, for both tasks, that an increase in AR alignment will decrease the probability of overriding an AI recommendation, supporting H 3. Additionally, providing global feature importance increases the likelihood of overriding recommendations for task I. Perceiving the model as fairer decreases the likelihood of overriding an AI recommendation for task II. Alignment in distributive fairness has no significant impact on both tasks. Additionally, we find evidence for the influence of global explanations (H 4) and fairness perception (H 5).

### 5.3 Gender Bias Alignment and Outcome

We analyze the influence on the team's outcome, focusing not on the team's distributive fairness but on the difference between the final human decision and the AI recommendation. Our results indicate that alignment in distributive fairness significantly influences the difference between the AI's and the teams' distributive fairness, supporting H 6. With increasing alignment, the difference decreases. Additionally, fairness perception has a significant effect in the same direction (see Table 3).

Table 2: Comparison of beta regression results with standard errors (SE) for overriding prevalence in both tasks: credit scoring vs. parole decision. This table presents the coefficients and SEs of predictors in both tasks, indicating their effect on overriding.

| *Depended Variable: Overriding Prevalence* | | | | |
|---|---|---|---|---|
| | **Credit Scoring** | | **Parole Decision** | |
| | Coeff. | SE | Coeff. | SE |
| Constant | 0.55 | (0.745) | 3.89*** | (0.802) |
| Fairness Perception | -0.05 | (0.110) | -0.39** | (0.128) |
| Alignment: Distributive Fairness | 0.51 | (0.857) | -0.48 | (0.610) |
| Alignment: Acceptance Rates | -3.58*** | (0.805) | -5.34*** | (0.884) |
| Explanations Provided | 0.68*** | (0.184) | 0.02 | (0.091) |
| Observations | 134 | | 132 | |

*Note:* *p<0.05; **p<0.01; ***p<0.001

## 6 Discussion

In our experiment, we investigate the role of gender bias alignment between humans and AI in human-AI teams. Specifically, we examine the influence of gender bias alignment on fairness perceptions (RQ 1), reliance (RQ 2), and the human-AI team's distributive fairness (RQ 3). We further analyze the role of explanations by providing global feature importance. Two experiments are carried out in which participants are asked to complete tasks in high-stakes contexts. In the first experiment, participants assess the credit scores of applicants (task I). In the second one, participants assess whether convicts would reoffend (task II). Our findings establish gender bias alignment as an important factor in human-AI team dynamics, which is strongly related to confirmation bias. The more gender biases of AI and humans align, the fairer the AI system is perceived, and the fewer recommendations are overridden. Additionally, gender bias alignment is pivotal for the team's distributive fairness. If biases are misaligned, the team's distributive fairness mainly hinges on human discretion, whereas the AI system's influence diminishes.

### 6.1 Theoretical Contributions

We demonstrate that gender bias alignment between humans and AI directly influences humans' perception of AI fairness. Our study extends previous research on confirmation bias [3, 7, 8, 89], emphasizing that humans tend to selectively adhere to decisions that are in harmony with their own beliefs. Also, such harmony directly influences their perception. Our findings highlight that the effect exists not only for expert knowledge and explicit expectations of AI recommendations but also for more subtle expectations like gender bias. In this sense, our study indicates that humans are sensitive to alignment, even if not explicitly stated and only implicitly presented through the AI recommendations. A possible explanation of our findings can be drawn from psychology. The belief-disconfirmation paradigm in dissonance research describes situations in which people encounter information that contradicts their beliefs. If they are unable or unwilling to reduce the resulting dissonance by changing their belief, they may respond by misperceiving, misinterpreting, rejecting, or refuting the conflicting information [30, 36]. Logically, when gender bias is misaligned, perceiving the AI as unfair may help reduce dissonance. Similarly, overriding recommendations can serve as a mechanism to alleviate cognitive dissonance in cases

Table 3: Comparison of regression results with standard errors (SE) for team fairness in both tasks: credit scoring vs. parole decision. This table presents the coefficients and SEs of predictors in both tasks, indicating their effect on fairness perception.

| *Dependent Variable: Difference in Fairness (AI vs. Final Human Decision)* | | | | |
|---|---|---|---|---|
| | **Credit Scoring** | | **Parole Decision** | |
| | Coeff. | SE | Coeff. | SE |
| Constant | 1.320*** | (0.164) | 0.321 | (0.270) |
| Fairness Perception | -0.049* | (0.024) | 0.013 | (0.040) |
| Alignment: Distributive Fairness | -0.994*** | (0.189) | -0.646*** | (0.162) |
| Alignment: Acceptance Rates | -0.263 | (0.176) | -0.324 | (0.296) |
| Explanations Provided | 0.083* | (0.040) | -0.056 | (0.039) |
| Observations | 134 | | 132 | |

*Note:* *p<0.05; **p<0.01; ***p<0.001

Table 4: Results summary for research questions (RQs) and hypotheses (H) for both tasks, with "X" marking significance.

| Research Question | Hypotheses | Factor | Credit Scoring | Parole Decision |
|---|---|---|---|---|
| **RQ1** | H1 | Alignment: Distributive Fairness | - | X |
| | H1 | Alignment: Acceptance Rate | X | X |
| | H2 | Explanations Provided | - | - |
| **RQ2** | H3 | Alignment: Distributive Fairness | - | - |
| | H3 | Alignment: Acceptance Rate | X | X |
| | H4 | Explanations Provided | X | - |
| | H5 | Fairness Perception | - | X |
| **RQ3** | H6 | Alignment: Distributive Fairness | X | X |
| | H6 | Alignment: Acceptance Rate | - | - |

of misalignment. Regarding fairness, our work provides a further connection between humans' perception of AI fairness and overriding prevalence, supporting previous results [7, 70]. Our findings additionally show an intriguing relationship between gender bias alignment and distributive fairness of the human-AI team. When biases are misaligned, humans tend to override AI recommendations more often. As a result, the final decision is increasingly converging towards the initial human decision without AI input, mirroring the human's "raw" distributive fairness.

## 6.2 Practical Implications

Besides theoretical implications, our results have important implications for effectively designing fair human-AI teams. Understanding which factors influence fairness perceptions and, by extension, influence reliance is crucial. Our results indicate that humans tend to view AI as fair when it echoes their own biases rather than when it conforms to a given standard (such as demographic parity). Consequently, it is essential for developers, deployers, and regulators to understand that AI fairness in human-AI teams involves more than embedding distributive fairness in AI recommendations, as it is a common practice in fair ML; it also requires accounting for the human-in-the-loop that makes the final decision [8, 66, 84, 89]. This is particularly relevant in light of recent EU regulations in the form of the AI Act [27] and the Ethics guidelines for trustworthy AI [28] that emphasize the role of human oversight in high-risk AI systems.

Prior work has already challenged predominant optimism towards human oversight, particularly concerning fairness [47]. When AI systems are supposed to follow concrete fairness objectives (such as demographic parity), human oversight can be detrimental to these objectives—and feature-based explanations might even reinforce this effect [70]. As a consequence, practitioners should define specific fairness objectives, establish monitoring mechanisms, and analyze possibilities to *effectively* enforce these objectives [20, 48]. Sterz et al. [83] identify four criteria for effective human oversight: causal power, epistemic access, self-control, and fitting intentions. Fulfilling all of these criteria is an ongoing challenge, necessitating the development of novel mechanisms (e.g., training programs, nudging strategies, or visual interfaces) and further empirical investigation to evaluate their effectiveness in specific applications. One possibility to account for these challenges is not using AI to make (potentially dissonant) recommendations but to provide nuanced evidence for or against decisions in the form of "evaluative AI" [58].

## 6.3 Limitations and Future Work

Like any empirical study, ours also has its limitations. For example, alignment is assessed solely in terms of acceptance rates and distributive fairness w.r.t to gender. Bias or fairness can be defined and measured in various ways, and it remains to be seen how the alignment with different biases and measurements would affect the results. We particularly leave the role of ground truth labels and fairness-accuracy trade-offs [32, 88] for future work. To make our results more generalizable and to fully understand the relationship between bias alignment and fairness perception, analyzing bias concerning other sensitive attributes such as age or ethnicity would be necessary as well. Additionally, our participants are exclusively from the USA and predominantly identify as White or Caucasian, which limits the scope for considering the impact of differing socialization experiences in non-Western contexts. The same applies to gender diversity, as we have only considered binary genders for simplicity. Future work could benefit from exploring a broader range of gender identities. Moreover, our task design comes with its limitations. For task I, the selection of income as a proxy for eligibility might have uncontrolled effects, as income is a measurable fact, whereas eligibility is a normative or economic decision. Further, credit eligibility is also influenced by factors like expected income changes, inferred from attributes such as age or occupation. These factors impact human decisions but are not reflected in the ground truth label. Moreover, the dataset, sourced from 1994, reflects a median household income of $26,000 [23], with incomes above $50,000 being uncommon compared to 2023's median of $80,000 [11]. Additionally, the self-reported perceived fairness for task II is highly skewed, with almost 80% of participants perceiving the AI as fair. We note that our participants are primarily laypeople, leaving open questions about how domain experts perceive and rely on AI. Future work could also investigate how biases among affected parties, rather than decision-makers, relate to fairness perceptions of AI systems. Finally, the advice provided by the AI has highly limited informational value, as we only examine the effect of recommendations and global feature importance. Future work might explore the effect of different means of information.

## 7 Conclusion

AI-informed decision-making is highly susceptible to replicating and amplifying social inequalities, driven by historical data biases embedded in machine learning models as well as individual biases introduced by human decision-makers. While much focus has been placed on algorithmic fairness to mitigate these effects in high-stakes situations, decisions are usually not made but rather supported by AI, prompting research on fairness dynamics emerging from human-AI collaboration. In our study, we explore the impact of gender bias alignment on decision-makers' fairness perceptions and reliance. We discover that fairness perceptions are not necessarily dependent on the presence of gender bias in AI systems but rather on whether the AI's level of gender bias aligns with the human level of gender bias. As our analysis of reliance behavior shows, this effect also translates into the final decisions, ultimately affecting the distributive fairness of human-AI team decisions.

Overall, our findings offer a novel perspective on reliance behavior by considering the particular alignment of human and AI characteristics in the form of gender bias. We extend the application of confirmation bias in the context of human-AI collaboration and offer fruitful pathways for future research. For example, future work could explore the role of alignments of other fairness and performance metrics. In particular, future work could pick up discussions of fairness-accuracy trade-offs and how they relate to the concept of bias alignment.

Finally, our work has significant implications for the application of fairness-enhancing techniques. As soon as a human-in-the-loop is installed, merely constructing a "formally fair" AI is insufficient to enforce fairness objectives; AI recommendations will likely be overridden if biases do not align. If human-AI teams are supposed to comply with these fairness objectives, we need effective methods to overcome misalignment and resolve ethical discrepancies.

# A Detailed Overview of Formulas for the Measurement Concept

At this point, we will give a more comprehensive overview and description of the measurement concepts. We provide this information as a reference for the interested reader and for reproducibility purposes.

*Measuring confirmation bias.* In this study, we explore confirmation bias beyond superficial alignment with expertise, focusing instead on value alignment in behavior, particularly gender bias—we call this gender bias alignment. We measure gender bias alignment using two metrics: acceptance rate (AR) and distributive fairness based on demographic parity (DP).

AR, defined in Equation (A.1), represents the conditional probability of an individual being accepted based on their gender and the decision-maker ($DM \in \{AI, human(H)\}$), where $gender \in \{men(M), women(W)\}$. Here, we focus on a binary decision where one is associated with a positive outcome (e.g., accepted application) or a negative outcome (e.g., denied application). Hence, the decision $\hat{Y}$ can be either accepted or denied. This metric analyzes how acceptance rates vary by decision-maker and gender, highlighting potential biases or preferences in processes like recruitment or admissions.

$$AR(DM, gender) = P[\hat{Y} = accepted|DM, gender], \\ gender \in M, W, DM \in AI, H \quad (A.1)$$

Based on the definition of AR, we quantify AR alignment as the sum of absolute differences in acceptance rates across genders. Higher values indicate greater alignment, while lower values indicate misalignment. The values are normalized to the interval $[0, 1]$ to ensure consistency and comparability. The formal definition of AR alignment is provided in Equation (A.2).

$$Alignment(AR) = 0.5 * (2 - (|AR(AI, M) - AR(H, M)| \\ + |AR(AI, W) - AR(H, M)|)), \\ AL \in [0, 1] \quad (A.2)$$

Equation (A.3) defines DP as the difference in the probabilities of a positive outcome ($\hat{Y} = accepted$) for men and women, conditioned on the decision-maker (DM). A value of zero indicates no bias, while positive values reflect a bias towards men. This metric quantifies and assesses gender bias in decision-making processes.

$$DP(DM) = P[\hat{Y} = accepted|M, DM] \\ - P[\hat{Y} = accepted|W, DM] \quad (A.3)$$

DP alignment is defined as the absolute difference between $DP(AI)$ and $DP(H)$. Higher values indicate greater alignment and are normalized to the interval $[0, 1]$ for consistency and comparability. The formal definition is provided in Equation (A.4).

$$Alignment(DP) = 0.5 * (2 - |DP(AI) - DP(H)|), \\ AL \in [0, 1] \quad (A.4)$$

*Measuring reliance.* Reliance captures how humans respond to AI recommendations, encompassing automation bias [3, 18, 55, 69], algorithmic aversion [22], and appropriate reliance [50, 68]. These archetypes describe whether humans override or adhere to AI recommendations. Consistent with prior studies [70], we define reliance as the probability of overriding an AI recommendation (see Equation (A.5)).

$$P(overriding) = \frac{\#overridenAIrecommendations}{\#decisions} \quad (A.5)$$

*Measuring difference in distributive fairness.* Distributive fairness is quantified using metrics that translate social values into computational terms [57]. However, such metrics are only weak approximations of the complex concept of fairness [59] and often involve trade-offs due to their incompatibility [31, 43]. Despite these limitations, demographic parity is suitable for our study due to its simplicity and its acceptance among laypeople [80]. We aim to examine the relationship between gender bias alignment, the team's distributive fairness—measured by the distributive fairness of the final human decisions—and the distributive fairness of the AI systems' recommendations. To achieve this, we quantify the difference in distributive fairness as the disparity in demographic parity (DP) between the AI recommendations ($DP(AI)$) and the final human decisions ($DP(H_{final})$) (see Equation (A.6)).

$$DP_{diff} = DP(H_{final}) - DP(AI) \quad (A.6)$$

# B Detailed Description for Task II: Parol Officer

Below we will illustrate the pre-processing, AI model development, and generation of explanations for task II.

## B.1 Pre-processing, AI Development, and Explanations

We follow a similar process to pre-process the COMPAS dataset, train two AI models, and extract global explanations. For feature selection, we base our approach on [16], excluding the description of the current charge. The final dataset includes the following features (ordered alphabetically): Age, Degree of Current Charge, Felony Crime Count before Age 18, Gender, Misdemeanor Crime Count before Age 18, and Prior Crime Count after Age 18. To induce or reduce biases, we undersample convicts who identify as men who reoffend and who do not, respectively, creating a gendered and ungendered dataset. The ungendered model achieves an accuracy of 85% and a distributive fairness of 0.02, indicating minimal gender bias. The gendered model achieves an accuracy of 89% and a distributive fairness of 0.7, indicating significant gender bias. An exemplary process is illustrated in Figure B.1

## B.2 Sample Generation

We follow the same workflow as applied to the COMPAS dataset, but provide additional material to foster the understanding. First, the characteristics of the ten samples were defined, ensuring an 80% accuracy for both AI models to include all types of errors. Gender bias, measured by Disparate Parity (DP), was established for each AI model. The ungendered AI demonstrated zero parity with a 40% Acceptance Rate (AR), while the gendered AI showed a 0.4 parity, with ARs of 60% for men and 20% for women. Both datasets maintained a 50:50 men-to-women ratio. For the ungendered AI, the samples were divided into eight groups based on gender (men/women), ground truth label (reoffence/no-reoffence), and the AI's recommendation (reoffence/no-reoffence). Group frequencies were determined by the accuracy and parity settings and are represented in Table B.1. The respective number of samples is then randomly drawn from each of the groups in the ungendered dataset and matched with the corresponding groups in the gendered AI dataset to ensure comparability while preserving randomness. To facilitate calculations of accuracy and DP, the respective confusion matrices are provided in Tables B.2 and B.3 for the parol officer task (Task II).

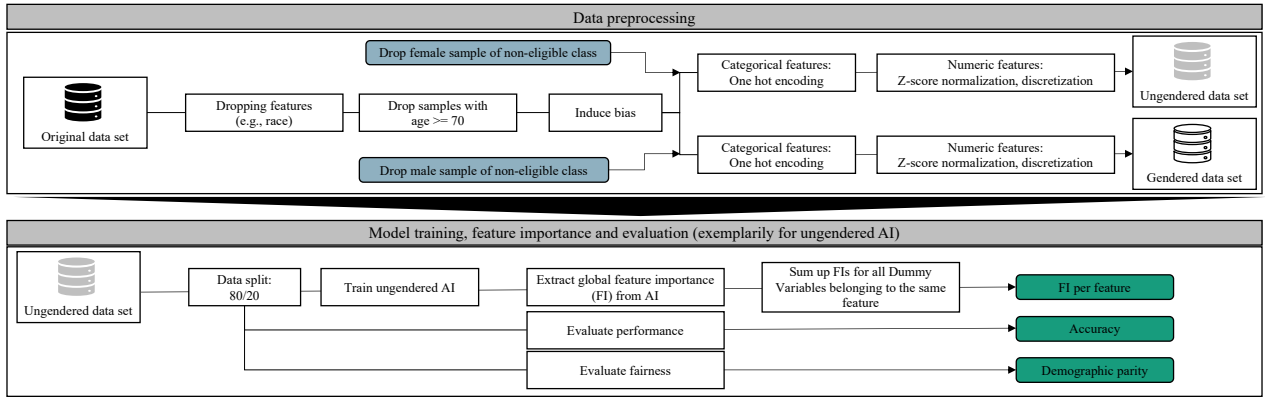

Figure B.1: Workflow of pre-processing, AI mode development, and explanations. The figures present an overview of the process, starting from the original dataset to the ungendered dataset and the training of the AI model. The example illustrates the complete workflow for the ungendered version of Task I (credit approval).

Table B.2: Confusion matrices for the ungendered datasets (men, women, and total).

| Ungendered - Men | | |
|---|---|---|
| Ground Truth \AI Decision | Reoffence | No Reoffence |
| Reoffence | 2 | 1 |
| No Reoffence | 0 | 2 |

| Ungendered - Women | | |
|---|---|---|
| Ground Truth \AI Decision | Reoffence | No Reoffence |
| Reoffence | 1 | 0 |
| No Reoffence | 1 | 3 |

| Ungendered - Total | | |
|---|---|---|
| Ground Truth \AI Decision | Reoffence | No Reoffence |
| Reoffence | 3 | 1 |
| No Reoffence | 1 | 5 |

Table B.3: Confusion matrices for the gendered datasets (men, women, and total).

| Gendered - Men | | |
|---|---|---|
| Ground Truth \AI Decision | Reoffence | No Reoffence |
| Reoffence | 2 | 0 |
| No Reoffence | 1 | 2 |

| Gendered - Women | | |
|---|---|---|
| Ground Truth \AI Decision | Reoffence | No Reoffence |
| Reoffence | 1 | 1 |
| No Reoffence | 0 | 3 |

| Gendered - Total | | |
|---|---|---|
| Ground Truth \AI Decision | Reoffence | No Reoffence |
| Reoffence | 3 | 1 |
| No Reoffence | 1 | 5 |

Table B.1: Dataset comparison of AI decision-making based on gender and ground truth labels.

| Dataset | Gender | Ground truth label | AI Decision | # Samples |
|---|---|---|---|---|
| ungendered | Men | reoffence | reoffence | 2 |
| | Men | reoffence | no reoffence | 1 |
| | Men | no reoffence | reoffence | 0 |
| | Men | no reoffence | no reoffence | 2 |
| | Women | reoffence | reoffence | 1 |
| | Women | reoffence | no reoffence | 0 |
| | Women | no reoffence | reoffence | 1 |
| | Women | no reoffence | no reoffence | 3 |
| gendered | Men | reoffence | reoffence | 2 |
| | Men | reoffence | no reoffence | 0 |
| | Men | no reoffence | reoffence | 1 |
| | Men | no reoffence | no reoffence | 2 |
| | Women | reoffence | reoffence | 1 |
| | Women | reoffence | no reoffence | 1 |
| | Women | no reoffence | reoffence | 0 |
| | Women | no reoffence | no reoffence | 3 |